\documentclass[12pt,letterpaper]{article}
\usepackage[a4paper, total={7in, 10in}]{geometry}

\usepackage{graphicx}
\usepackage{helvet}
\usepackage{authblk}
\usepackage{hyperref}
\usepackage{amsmath} 
\usepackage{amssymb} 
\usepackage{orcidlink} 
\usepackage[super,comma,sort&compress]  
   {natbib}

\makeatletter

\renewcommand{\maketitle}{\bgroup\setlength{\parindent}{0pt}
\begin{flushleft}
  \textbf{\@title}
  
  \@author
\end{flushleft}\egroup}
\makeatother

\title{Optical Resonances: From Eigenmodes to Scattering Features}
\date{}

\author[1]{Ilya Karavaev}
\author[2,*]{Kirill Koshelev}
\author[1,*]{Andrey Bogdanov}

\affil[1]{Qingdao Innovation and Development Center of Harbin Engineering University, Qingdao 266000, Shandong, China}

\affil[2]{Department of Electronic Materials Engineering, Research School of Physics,
Australian National University, Canberra, ACT, Australia}

\affil[*]{Correspondence: kirill.koshelev@anu.edu.au (K.K.), a.bogdanov@hrbeu.edu.cn (A.B.)}

\begin{document}

\maketitle

\section*{SUMMARY}

Electromagnetic resonances play a central role in nanophotonics by enabling efficient confinement of electromagnetic energy and enhanced light-matter interaction. Traditionally, resonant phenomena have been described using platform-specific concepts developed within distinct research communities, including photonic crystals, plasmonics, and dielectric metasurfaces. In this Perspective, we propose a unified framework that distinguishes electromagnetic resonances as eigenmodes of open systems from their experimentally observed manifestations as scattering features. We show how resonances evolve from isolated particles to coupled oligomers and periodic structures, highlighting the roles of geometry, material response, and dimensionality. Particular attention is given to interference-driven phenomena such as bound states in the continuum, lattice resonances, anapoles, and superscattering, some of which cannot always be associated with a single eigenmode. By clarifying the relationship between eigenmodes, scattering channels, and interference effects, this Perspective provides a coherent language for interpreting resonant phenomena and identifies key challenges and opportunities for designing robust resonant photonic systems.

\section*{KEYWORDS}

photonic crystal, plasmonics, high-contrast grating, metasurface, bound states in the continuum, Fano resonance, lattice resonance, anapole, scattering, quasi-normal modes

\section*{INTRODUCTION}

Electromagnetic resonances are a cornerstone of modern optics and photonics, governing light-matter interaction, spectral selectivity, and scattering responses across a wide range of systems. By confining electromagnetic energy, resonant structures substantially enhance the light-matter interaction. As a result, resonant effects underpin essential optical functionalities, including sensing, nonlinear frequency conversion, and lasing, and form the basis of many photonic devices and meta-structures~\cite{bothResonantStatesTheir2021,koshelevDielectricResonantMetaphotonics2021}.

Despite their fundamental importance, resonant phenomena are commonly described using platform-specific terminology that emerged from historically distinct research communities. Photonic crystals, plasmonic nanostructures, high-contrast gratings, and dielectric metasurfaces each emphasize different physical aspects, such as band formation, material dispersion, or geometric confinement, resulting in parallel but fragmented descriptions of closely related effects. Resonant features, including dark and bright modes~\cite{yang2010plasmon}, Fano interference~\cite{luk2010fano}, guided-mode resonances~\cite{magnusson2016guided}, lattice resonances~\cite{evlyukhin2010optical}, and bound states in the continuum (BICs)~\cite{koshelev2023bound}, frequently appear across these platforms, yet are often treated as fundamentally different phenomena. This conceptual fragmentation is not merely semantic. It directly affects how resonances are designed, measured, and understood in experiments. At a fundamental level, many of these effects stem from the same physical mechanisms, but this common origin is often obscured when resonances are classified by assigning them to a specific material platform. This situation is further complicated when optical resonances are mixed with their manifestations, namely various scattering features such as Fano resonances, Wood anomalies, or electromagnetically induced transparency.

In this Perspective, we propose a unified viewpoint on electromagnetic resonances and scattering features based on system dimensionality, material response, and their relation to non-Hermitian eigenmodes. Rather than providing an exhaustive survey, we aim to clarify shared physical principles, connect theoretical descriptions with experimental observables, and highlight key challenges in the practical realization of resonant photonic systems.

\section*{HISTORICAL OVERVIEW}

Optical resonances have been studied for more than a century, however, their importance increased dramatically following the invention of the laser in the early 1960s, when optical cavities based on Fabry--P\'{e}rot resonators enabled coherent light amplification and mode selection~\cite{maimanStimulatedOpticalRadiation1960}. The role of resonances became even more prominent with the subsequent development of distributed-feedback and surface-emitting lasers in the 1970s, laying the foundation for modern coherent light sources~\cite{kogelnikStimulatedEmissionPeriodic1971, DFB_suris}. 

During the 1990s and 2000s, the focus shifted to periodic semiconductor structures, mostly driven by the development of photonic crystal slabs~\cite{sakodaOpticalTransmittanceTwodimensional1995}. These systems supported guided and radiative resonances, enabling planar on-chip control of light and advancing integrated photonics~\cite{tikhodeevQuasiguidedModesOptical2002}. Their study contributed fundamentally to the understanding of diffraction, photonic band formation, and field localization in periodic media. In the 2000s, attention turned to plasmonic systems, in which collective oscillations of free electrons in metals provided a mechanism for confining light below the diffraction limit~\cite{maierPlasmonicsRouteNanoscale2001}. Plasmonic resonances enabled the development of nanoscale optical antennas, sensors, ultrathin waveguides,  and various field-enhancement platforms. Plasmonic waveguides enabled efficient terahertz laser architectures and played an important role in the commercialization of modern THz laser sources~\cite{williamsTerahertzQuantumcascadeLaser2003}. Despite significant progress, intrinsic absorption losses in metals posed a major limitation for selected practical applications~\cite{khurginHowDealLoss2015}, motivating the search for alternative low-loss resonant architectures~\cite{koshelevDielectricResonantMetaphotonics2021}. 

From the late 2000s onwards, high-index dielectric nanophotonics, including high-contrast gratings and individual resonators, emerged as a promising low-loss alternative to plasmonic and bulk metamaterial platforms~\cite{lalanneOpticalPropertiesDeep2006}. Non-magnetic dielectric nanoparticles and metasurfaces were shown to support electric and magnetic Mie-type resonances with negligible absorption, enabling field localization and scattering control comparable to, or even surpassing, plasmonic systems~\cite{kuznetsovMagneticLight2012}. This rapidly developing direction, often referred to as {\it Mie-tronics}~\cite{wonMietronicEra2019}, established a new paradigm for resonant light control based on dielectric building blocks. Recently, the focus has shifted toward more complex dielectric and hybrid metal-dielectric structures that can simultaneously provide high quality factors and small mode volumes.~\cite{ouyang2024singular} 

Throughout this historical progression, similar resonant phenomena, such as Fano interference, dark- and bright-mode formation, and hybridization between localized and extended modes, have repeatedly appeared across different physical settings. These effects were rediscovered across lasers, photonic crystals, plasmonic nanostructures, and dielectric metasurfaces as fabrication capabilities and operating frequencies evolved. While each platform emphasized different physical realizations, the recurring appearance of analogous resonant behavior reflects common underlying mechanisms that transcend specific material systems or geometries.

\section*{FROM LOCAL CONFINEMENT TO COLLECTIVE MODES}

\subsection*{GEOMETRIC VERSUS MATERIAL RESONANCES}

Optical resonances can be broadly divided according to their physical origin into \textit{geometric} and \textit{material} resonances, as illustrated in Fig.~1. 
Geometric resonances arise from electromagnetic field confinement due to refractive-index contrast between a structure and its surrounding environment. Their characteristic scale is determined by the number of wavelengths that fit inside the resonator. Well-known examples include Mie resonances in dielectric particles, Fabry-P\'erot (FP) resonances, whispering-gallery modes (WGMs), and the modes of photonic crystal cavities. Material resonances, by contrast, originate from intrinsic excitations of the medium itself, such as collective electronic oscillations, excitons, or optical phonons. These excitations can strongly couple to the electromagnetic field, giving rise to hybrid light-matter states known as {\it polaritons}. The most prominent example is the plasmonic resonance, whose frequency is primarily determined by material parameters such as the free-carrier concentration and effective mass. In this case, geometry mainly governs the strength of coupling to radiation and the resulting linewidth, rather than setting the resonance frequency itself.

\subsection*{WHEN A SINGLE RESONATOR IS ENOUGH}

Single resonators are the simplest class of resonant systems shown in Fig.~1.  In high-index dielectric particles, Mie-type resonances form the foundation of dielectric nanophotonics. Magnetic-type resonances arise naturally when polarization currents form circulating vortex patterns, thereby generating an effective magnetic dipole moment. In contrast to small particles in the Rayleigh limit, which are commonly described as electric dipole scatterers, the fundamental resonant mode of high-index dielectric particles is the magnetic dipole resonance. While Mie theory was originally formulated for spherical particles~\cite{mie1908beitrage}, the same physical picture extends to resonators of arbitrary shapes, where each eigenmode can be described by a multipolar expansion, typically dominated by a single multipole contribution. Suppressing radiation associated with this dominant multipole through symmetry or geometric tuning leads to a drastic increase in the Q factor and a qualitative modification of the scattering response. This regime is commonly referred to as a \textit{supercavity mode} or a \textit{quasi-bound state in the continuum} in an isolated resonator, and has enabled applications such as nanolasing and enhanced nonlinear light conversion~\cite{zalogina2023high}. Very recently, it was shown that Mie resonances can be efficiently utilized in inverse cavities called \textit{Mie voids}, where the light is confined in air surrounded by a thick high-index dielectric substrate~\cite{hentschel2023dielectric}. Mie voids were shown to extend the range of applicability of the Babinet principle~\cite{hamidi2025quasi}, proving a single-resonator platform for the enhancement of quantum and nonlinear effects~\cite{lu2025light}.

Axially symmetric dielectric resonators, such as disks, rings, and toroids, can also support WGMs, in which light circulates along the resonator's perimeter. Radiative losses of these modes decrease exponentially with resonator size, enabling exceptionally high quality factors, often exceeding $10^{10}$~\cite{savchenkovOpticalResonatorsTen2007}. Due to their weak coupling to free-space radiation, WGMs are typically excited using evanescent coupling via prisms, waveguides, or tapered fibers~\cite{foreman2016dielectric}. 

Plasmonic nanoparticles support pronounced localized surface plasmon resonances in the visible spectral range despite their deeply subwavelength dimensions. Although intrinsic metallic losses limit the quality factor of these resonances, their extremely small mode volumes result in strong electromagnetic field confinement and giant local field enhancement, which are crucial for applications such as biosensing and photocatalysis~\cite{mayerLocalizedSurfacePlasmon2011}. Plasmonic materials exhibit a distinctive epsilon-near-zero (ENZ) regime at frequencies close to the plasma frequency. In this regime, electromagnetic waves can be squeezed and tunneled through deeply subwavelength channels filled with ENZ material~\cite{ENZ-Engheta}, offering strong potential for nonlinear nanophotonic applications~\cite{reshefNonlinearOpticalEffects2019}. Moreover, ENZ materials can act as electromagnetic shields in open resonators, suppressing radiation loss and enabling true BICs in individual structures~\cite{silveirinhaTrappingLightOpen2014}. 

\subsection*{EMERGENCE THROUGH COUPLING: DIMERS AND OLIGOMERS}

Coupling individual resonators into dimers or larger oligomers gives rise to hybridization and interference effects, as indicated in the central part of Fig.~1. In particular, hybridization of resonant modes in dimers results in the formation of bright and dark states. Bright modes couple efficiently to the far-field and dominate the scattering response, whereas dark modes exhibit suppressed radiation due to symmetry or accidental destructive interference, enabling higher quality factors and enhanced energy confinement. These effects occur in both dielectric and plasmonic oligomers and provide a flexible route for tailoring resonance strength, linewidth, and radiation patterns. Near-field interaction between closely spaced particles leads to the formation of gap modes, in which electromagnetic energy is concentrated into deeply subwavelength hot spots~\cite{lei2012revealing}. Such gap modes can play a central role in plasmonic field enhancement, sensing, and nonlinear optics. The extreme case of a gap mode is realized in sub-nanometer metallic gaps, where atomic-scale features form picocavities with unprecedented field confinement.

Oligomers formed by several coupled nanoparticles provide additional degrees of freedom for engineering resonance spectra through controlled symmetry, inter-particle coupling, and modal interference~\cite{chong2014observation}. In such systems, as in dimers, hybridization of individual particle modes gives rise to bright and dark collective states, whose interference produces pronounced, spectrally tunable Fano resonances. In plasmonic oligomers, collective coupling can give rise to effective magnetic dipole responses absent in individual particles, resulting in pronounced bianisotropy and enabling directional scattering effects, such as the Kerker effect. 

\subsection*{WHY PERIODICITY CHANGES EVERYTHING}

Photonic structures with continuous or discrete translation symmetry form a distinct class of resonant systems, as summarized in the bottom part of Fig.~1. Unlike isolated resonators, they confine electromagnetic fields by suppressing radiation away from the interface, slab, or waveguide. This suppression can originate from total internal reflection, reflection from media with negative permittivity or a Bragg mirror, surface wave confinement, or polarization mismatch between guided states and external transverse waves. The latter mechanism is relevant to ENZ and plasmonic media, where longitudinal plasma-like excitations are decoupled from propagating electromagnetic modes. 

Guided modes are truly bound states with dispersion below the light line and, therefore, do not couple directly to free-space radiation. They can nevertheless be excited externally through frustrated TIR, grating couplers, structural defects, or nonlinear processes that provide the required momentum matching. By contrast, FP modes, also known as leaky modes, are not truly bound because the TIR condition is not fulfilled. As a result, they radiate into the surrounding space and can be directly accessed from the far field at normal or oblique incidence.

Surface states can exist at interfaces separating media with opposite signs of the real parts of their permittivities, i.e., $\mathrm{Re}(\varepsilon_1)\mathrm{Re}(\varepsilon_2)<0$. This condition is realized at metal--dielectric interfaces, giving rise to surface plasmon polaritons (SPPs), and at polar-crystal--dielectric interfaces within the Reststrahlen band, where surface phonon polaritons can form. Related interface states also include Tamm and Tamm--SPP modes, which are supported at the boundary between a distributed Bragg reflector and, respectively, a dielectric or metallic layer~\cite{kaliteevski2007tamm}. 

Periodic photonic structures, including metasurfaces, high-contrast gratings, corrugated waveguides, and photonic-crystal slabs, possess discrete translational symmetry. In addition to guided, FP, and surface modes, periodicity enables band folding modes and collective lattice responses. In particular, periodic modulation folds guided modes into the first Brillouin zone, forming quasi-guided modes (QGMs), or guided-mode resonances (GMRs)~\cite{tikhodeevQuasiguidedModesOptical2002}. These modes have finite radiative leakage and disappear as the periodic modulation vanishes, reflecting their hybrid guided-radiative nature. Unlike low-Q FP modes, which usually produce broad spectral modulations, QGMs give rise to sharp Fano features in reflection and transmission spectra~\cite{fanAnalysisGuidedResonances2002}.

Surface lattice resonances~\cite{kravetsPlasmonicSurfaceLattice2018a} are collective nonlocal excitations enabled by periodicity. They arise from the coupling of resonant unit-cell responses to diffractive lattice modes near Rayleigh anomalies. Unlike localized resonances, their electromagnetic energy is mainly stored in extended evanescent fields outside the structure, reducing overlap with lossy materials. This nonlocal character allows lattice resonances to reach comparatively high quality factors even in lossy plasmonic systems.

A particularly important and conceptually unifying phenomenon in periodic systems is the emergence of BICs~\cite{koshelev2023bound}. In periodic structures, radiation is allowed only into discrete diffraction channels. Remarkably, even when such channels are open, the corresponding radiation amplitude can vanish exactly, giving rise to non-radiating states embedded in the radiation continuum. This suppression of radiation can occur either due to symmetry constraints, leading to symmetry-protected BICs, or due to fine-tuning of structural parameters, resulting in so-called accidental BICs. In the momentum space, BICs manifest as polarization vortices in the far field carrying a topological charge.  

Experimentally, BICs are commonly accessed by intentionally breaking the symmetry of the structure or by varying the angle of incidence, which transforms them into nearby high-quality-factor resonances known as quasi-BICs~\cite{koshelev2018asymmetric}. This formalism turns out to be powerful enough to predict BICs even in the aperiodic structures, namely in a single dielectric ridge on the surface of the waveguide~\cite{bezus2018bound}. It is important to emphasize that BICs can exist even in plasmonic structures, despite intrinsic material losses. In this case, a BIC has a finite lifetime due to material losses while remaining fully decoupled from the radiation continuum. As a result, the total quality factor is finite, whereas the radiative quality factor diverges. Recently, BICs were observed in a plasmonic wired medium with strong spatial dispersion due to interference of longitudinal plasma-like waves and TEM-polarized modes~\cite{koreshin2025bound}.

\section*{EXPERIMENTAL OBSERVABILITY AND PRACTICAL LIMITS}

\subsection*{RESONANCES VERSUS SCATTERING FEATURES}

Optical systems are inherently open, and their resonances are therefore described by quasi-normal modes with finite radiative quality factors~\cite{bothResonantStatesTheir2021}. Unlike perfectly bound mathematical eigenstates, these modes lose energy through radiation and absorption, which makes them experimentally observable in the far-field. Importantly, eigenmodes are not measured directly. Instead, experiments probe the scattering response of the system to an external excitation. As a result, resonances manifest indirectly as spectral or temporal features in measured quantities such as reflection, transmission, absorption, or emitted radiation, summarized in Fig.~2.

While resonant eigenmodes provide the fundamental building blocks of the optical response, many prominent scattering phenomena are not associated with a single eigenmode. Instead, they arise from interference between different scattering channels or from the combined contribution of multiple resonant modes. A representative example is the anapole, which is not an eigenmode but a scattering condition characterized by suppressed radiation into a specific channel due to destructive multipolar interference~\cite{miroshnichenko2015nonradiating}. As a result, an anapole cannot be meaningfully described by a quality factor. In contrast, superscattering emerges when several resonant modes associated with different channels become spectrally aligned, leading to simultaneous enhancement of multiple partial cross-sections and an unusually large total scattering response~\cite{ruanSuperscatteringLightSubwavelength2010}. Related interference effects include the Kerker effect, where coherent electric and magnetic dipole radiation suppresses backscattering and enables directional response~\cite{kerker1983electromagnetic}. 

In periodic systems, Rayleigh anomalies mark the opening diffraction channels and, therefore, do not correspond to resonant eigenmodes. They typically appear as sharp kinks or discontinuities in the spectral response. By contrast, Wood anomalies are genuinely resonant features associated with the excitation of quasi-guided modes supported by the periodic structure. Their interference with the nonresonant background gives rise to characteristic Fano lineshapes in reflection or transmission~\cite{maradudin2016rayleigh}. Additional interference-driven phenomena include critical coupling~\cite{yariv2002critical}, coherent perfect absorption~\cite{baranov2017coherent}, and electromagnetically induced transparency~\cite{lukin2001controlling}, all of which result from a tailored balance or cancellation between radiative pathways and enable complete reflection suppression, enhanced absorption, or narrow transparency windows. 

Importantly, most of these scattering features in both single and periodic structures can be understood within the framework of Fano resonances, which provide a general description of reflection, transmission, scattering cross-sections, and even luminescence spectra~\cite{limonov2017fano}. However, the Fano resonance itself is not a distinct physical resonance of the system. Instead, it represents a universal spectral lineshape for scattering spectra, describing how one or more eigenmodes appear in the far field through their coupling to external scattering channels (see the bottom part of Fig.~2). In the simplest case of a single resonant mode, the Fano resonance is described by
\begin{equation}
\sigma = \frac{(q+\epsilon)^2}{1 + \epsilon^2},
\end{equation}
where $\epsilon = 2(\omega - \omega_\mathrm{res})/\gamma$, $\omega_\mathrm{res}$ is the resonant frequency and $q$ is the Fano asymmetry parameter. The total loss $\gamma=\gamma_\text{rad}+\gamma_\text{abs}$ is contributed by radiative and absorption losses. In realistic systems, measured spectra are typically governed by multiple or cascaded Fano resonances, reflecting the simultaneous contribution and interference of several modes.

\subsection*{EXCITATION PATHWAYS AND MEASUREMENT APPROACHES}

Experimental access to resonances in photonic systems generally relies on two complementary pathways, far-field and near-field excitation. In the far-field excitation scheme shown in Fig.~3a, the system is illuminated by an external light source, and the scattered, reflected, or transmitted intensity is measured as a function of frequency. From the measured spectra, one can extract the resonance frequency $\omega_\text{res}$ and linewidth $\gamma$, which quantify the combined radiative and non-radiative losses. Equivalent information can be obtained in the time domain, where pulsed excitation followed by time-resolved detection yields the modal lifetime~\cite{tanaka2007dynamic}. Spectral and temporal measurements are therefore complementary methods. Both far-field approaches enable the characterization of very high-Q resonances, with quality factors reaching $10^6$ and beyond~\cite{kodigala2017lasing}. 

Near-field excitation, illustrated in Fig.~3b, provides access to modes that are weakly radiating or entirely decoupled from the far-field. This includes modes below the light line, such as waveguide modes or SPPs, which require near-field or momentum-matching excitation. For high-$Q$ resonances, even small material absorption can place the system deep in the undercoupled regime ($\gamma_{\text{rad}}\ll\gamma_{\text{abs}}$), strongly suppressing the amplitude of the associated Fano feature and rendering such modes difficult or impossible to detect in conventional far-field measurements. As a result, WGMs and other weakly radiating resonances are most commonly accessed via near-field coupling, for example, through frustrated total internal reflection or by evanescent coupling to guided modes~\cite{righini2011whispering,odit2021observation}.

Near-field pumping further enables the excitation of modes via photoluminescence, lasing, or nonlinear optical signals, in which the induced polarization serves as an internal source of electromagnetic radiation. As shown schematically in Fig.~3c, the induced polarization currents may be incoherent, as in spontaneous emission or thermal radiation, or coherent, giving rise to phase-sensitive nonlinear processes such as sum-frequency generation. Thermal radiation and luminescence, therefore, constitute additional, inherently incoherent excitation mechanisms. Importantly, photonic resonances can strongly modify the local density of optical states, enabling substantial enhancement of emission and conventional or even polaritonic lasing.

\subsection*{FABRICATION IMPERFECTIONS: FINITE SIZE, DISORDER, ROUGHNESS}

Transitioning from idealized theoretical models to experimental setups exposes significant measurement challenges in the implementation of photonic devices based on specific resonant features~\cite{kuhne2021fabrication}. In theoretical treatments, perfect symmetry, infinite periodicity, and lossless materials are often assumed, leading to predictions of sharp, high-Q resonances with pronounced spectral signatures, such as perfect reflection or transmission, as illustrated by the dark black curve in Fig.~4a. In practice, however, real structures are necessarily finite and subject to material absorption, fabrication tolerances, surface roughness, size fluctuations, and structural disorder. These non-idealities collectively shift resonance frequencies, reduce quality factors, and weaken interference effects, as evidenced by the broadened and diminished response of the practical sample shown by the orange curve in Fig.~4a. Importantly, even minor deviations from ideal conditions can strongly affect high-Q resonances, emphasizing that experimentally measured spectra encode not only intrinsic modal properties but also finite-size effects and disorder. Addressing these challenges requires a dual strategy that combines continued advances in fabrication precision with the deliberate design of resonant structures that are intrinsically robust against imperfections and losses. At the same time, optimizing resonant behavior in finite-size photonic structures remains an active area of research, with various approaches aimed at minimizing device footprint while preserving high-Q resonances and strong field confinement~\cite{chenObservationMiniaturizedBound2022a,liuHigh$Q$QuasiboundStates2019a}. In parallel, considerable effort is directed toward developing photonic designs that maintain their resonant properties in the presence of structural disorder, including strategies based on BICs and concepts from topological photonics~\cite{jinTopologicallyEnabledUltrahighQ2019a,khanikaevTopologicalPhotonicsRobustness2024}.

\subsection*{MEASUREMENT BACK-ACTION: QUALITY FACTOR LOADING AND ANGULAR AVERAGING}

Beyond fabrication-related challenges discussed in previous sections, the experimental quantification of resonant properties faces fundamental limitations rooted in the nature of quasi-normal modes themselves. These modes are coupled to the radiation continuum and exhibit losses through both intrinsic dissipation ($\gamma_\text{abs}$) and radiative leakage ($\gamma_\text{rad}$). To detect such resonances experimentally, the mode must be efficiently excited and the total scattered field collected, often requiring the integration of additional elements, such as the waveguide illustrated in Fig.~4b, which optimizes signal extraction under specific phase-matching conditions. However, this coupling inherently modifies the system, and the measured loaded quality factor $Q_L$ becomes a composite quantity incorporating both the intrinsic losses of the resonance $(Q_i)$ and the external contributions from the measuring device $(Q_\text{ex})$, as expressed by the relation $1/Q_L = 1/Q_{i} + 1/Q_\text{ex}$~\cite{kajfez2003random}. Consequently, a key experimental challenge lies in distinguishing the intrinsic quality factor of the resonance from the loaded quality factor measured in the presence of the external coupling. In practice, this requires precise control over the coupling strength between the resonator and the measurement apparatus, for example, by adjusting the distance between a waveguide and the resonator. Without careful calibration or the use of decoupling techniques, the extracted Q-factor may significantly underestimate the actual performance of the resonant mode, obscuring its true potential for applications requiring high coherence or strong field confinement.

In addition, the experimental setup must be carefully designed to ensure selective excitation of the target mode while minimizing excitation of other resonances.  Resonant responses of periodic structures are typically described assuming excitation by a plane wave with a well-defined in-plane wave vector. Under such conditions, only modes matching this wave vector contribute to the measured reflection or transmission. In practice, however, experiments are typically performed with beams of finite aperture. Such beams usually represent a superposition of plane waves with slightly different propagation directions and in-plane wave vectors. As a result, multiple nearby resonances can be excited simultaneously. The measured reflected or transmitted intensity, therefore, corresponds to a weighted convolution of the responses associated with all angular components of the incident beam (Fig.~4c). This angular averaging leads to inhomogeneous broadening of the spectral features and complicates the extraction of the intrinsic properties of the underlying resonances.

A further complication arises from nonlinear and self-action effects that may occur during the measurement itself. In such cases, a careful interpretation of the experimental data requires a self-consistent analysis of the resonant response~\cite{sinevObservationUltrafastSelfAction2021}. High-Q resonances concentrate electromagnetic energy in small mode volumes, so even modest excitation powers can induce a nonlinear response. As a result, Kerr self-action, multiphoton absorption, thermo-optical shifts, and optical multistability may appear already at sub-milliwatt power levels~\cite{liuOpticalMultistabilityCompact2025}. These processes dynamically modify the resonance frequency, linewidth, and spectral lineshape during the measurement, thereby distorting the observed spectra and complicating the interpretation of the experimental data. 

\section*{OUTLOOK AND PERSPECTIVE}

Recent advances in nanophotonics and meta-optics have introduced new degrees of freedom for resonant photonics, including chirality, anisotropy, nonlinear response, and temporal modulation. Chiral resonant optics aims to engineer modes with controlled near- or far-field helicity for polarization-selective light control. Such responses can be realized through excitation geometry~\cite{ren2012giant}, complex meta-atom shapes~\cite{zhang2022chiral}, meta-atom rotation~\cite{toftul2024chiral,sinev2025chirality}, and twisted bilayer architectures~\cite{gromyko2024unidirectional}. Anisotropic media further enable unconventional resonant effects, including long-range hyperbolic polaritons~\cite{liu2025long}, anisotropy-induced BICs without periodicity~\cite{ovcharenko2020bound,gomis2017anisotropy,chen2024anisotropy}, and Janus BICs in shear-induced anisotropic metastructures~\cite{ji2026janus}. Another important direction is the reduction of the effective mode volume in dielectric structures beyond conventional diffraction limits~\cite{ouyang2024singular}. Recent dielectric platforms based on twisted cavities and bowtie-like geometries enable single-nanometer field localization with characteristic narwhal-shaped wavefunctions~\cite{mao2025singulonics,ma2026sub}. Such confinement enhances the electric field per photon and strengthens spontaneous emission, nonlinear response, and cavity-QED interactions~\cite{frisk2019ultrastrong,lu2021quantum}. In combination with high-Q resonances, nonlinear effects such as optical multistability, recently demonstrated in compact microcavities and photon-avalanching nanocrystals~\cite{liu2025optical,skripka2025intrinsic}, may enable low-power optical memory and all-optical switching. Time-varying media provide an additional route for controlling resonant eigenmodes and scattering features~\cite{galiffi2022photonics,caloz2019spacetime}. Temporal modulation enables frequency conversion, nonreciprocal transport, and dynamic control of radiative coupling~\cite{shaltout2015time,guo2019nonreciprocal}. In resonant systems, finite photon lifetime enhances even weak temporal perturbations, leading to active Fano lineshapes, tunable linewidths, sideband generation, and temporal programming of resonances~\cite{zhang2025photon}, naturally described within Floquet non-Hermitian modal frameworks~\cite{sun2025flatband}.

Resonant nanophotonics is increasingly moving from the discovery of fundamental effects, such as BICs, Fano interference, and topological photonic modes, toward the design of robust and scalable photonic devices. A key challenge is the sensitivity of high-Q resonances to the angular divergence of focused beams, which can broaden spectral features and degrade device performance, as discussed in Fig.~4c. Photonic flatbands offer a route to angularly robust resonant enhancement by maintaining low group velocity over an extended region of the Brillouin zone~\cite{leykam2018perspective,sun2025flatband}. Such flatbands can be engineered through Brillouin-zone folding~\cite{murai2022engineering,wang2023brillouin}; for example, period doubling in perturbed dimerized lattices folds additional bound states into the radiation continuum, giving rise to QGMs with controllable properties.

The integration of materials with unconventional symmetries is opening novel pathways for light control that extend far beyond the capabilities of traditional isotropic media~\cite{tselikov2025tunable}. While the techniques for nanostructuring these materials are rapidly maturing~\cite{munkhbat2023nanostructured}, the achievable precision and surface quality have yet to reach the rigorous benchmarks established by mature semiconductor platforms like Si, GaAs, or GaP. To overcome fabrication limitations such as surface roughness and material damage from etching, the concept of etchless metasurfaces has emerged~\cite{huang2023ultrahigh,chen2023high}. By patterning an auxiliary oxide or polymer layer instead of the primary high-index material, the structural and electronic properties of the active medium are preserved. 

Simultaneously, the focus is shifting from single-mode systems to structures exhibiting multi-resonant behavior~\cite{reshef2019multiresonant,zhou2023multiresonant,huang2024microcavity}. This multi-resonant approach provides a versatile route for broadband field enhancement and high-efficiency nonlinear optical processes. Physics-aware machine learning algorithms are another critical direction that allows for simplifying scalable metasurface patterning strategies~\cite{cheng2025closing}. By comparing a variety of structures of different configurations and the corresponding features of the spectrum, the method is able to effectively eliminate noise, compensate for distortions, and ultimately solve the inverse problem of finding a structure with a particular resonance at the desired frequency.

Looking ahead, the synergy of modal engineering, scattering-channel control, advanced material platforms, and high-quality large-scale fabrication may transform resonant nanophotonics into a general framework for robust, scalable, and multifunctional control of light--matter interaction, enabling next-generation technologies for optical computing, sensing, nonlinear optics, quantum photonics, and active light control.

\newpage

\section*{RESOURCE AVAILABILITY}

\subsection*{Lead contact}

Requests for further information and resources should be directed to and will be fulfilled by the lead contact, Kirill Koshelev (kirill.koshelev@anu.edu.au).

\subsection*{Data and code availability}

No new data were generated or analyzed in this study.

\section*{ACKNOWLEDGMENTS}

A.B. acknowledges support from the National Natural Science Foundation of China (Project W2532010).

\section*{AUTHOR CONTRIBUTIONS}

Conceptualization, K.K. and A.B.; methodology, I.K., K.K. and A.B.; writing-–original draft, I.K.; writing-–review \& editing, K.K. and A.B.; funding acquisition, A.B.; supervision, K.K. and A.B.

\section*{DECLARATION OF INTERESTS}

The authors declare no conflicts of interest.

\newpage

\section*{MAIN FIGURE TITLES AND LEGENDS}

\noindent\includegraphics[width=0.85\linewidth]{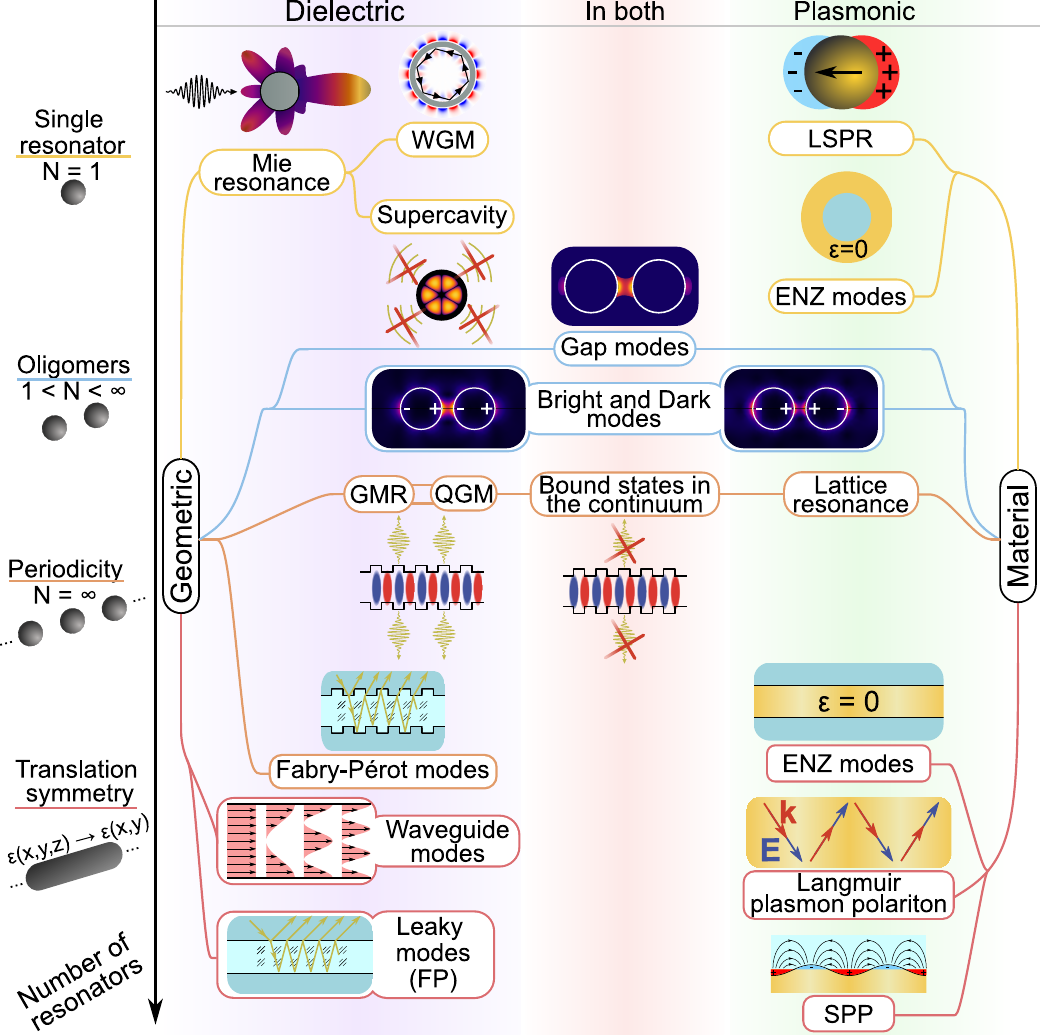}

\subsection*{Figure 1. Classification of electromagnetic resonances and scattering effects in dielectric and plasmonic systems, organized by dimensionality of the structure.}\label{fig1}

Classification of eigenmodes found in photonic systems with various configurations. The yellow lines highlight the effects in single resonators of arbitrary shape, the blue lines describe the ones in oligomers, the orange -- periodic structures, and the red -- the systems with translational symmetry. The features mentioned in the left column of the figure emerge mainly in dielectric structures and originate from the geometry of the resonator itself, while the right column describes phenomena based primarily on material parameters. The features that depend on both geometry and material are located in the central column.

\noindent\includegraphics[width=0.85\linewidth]{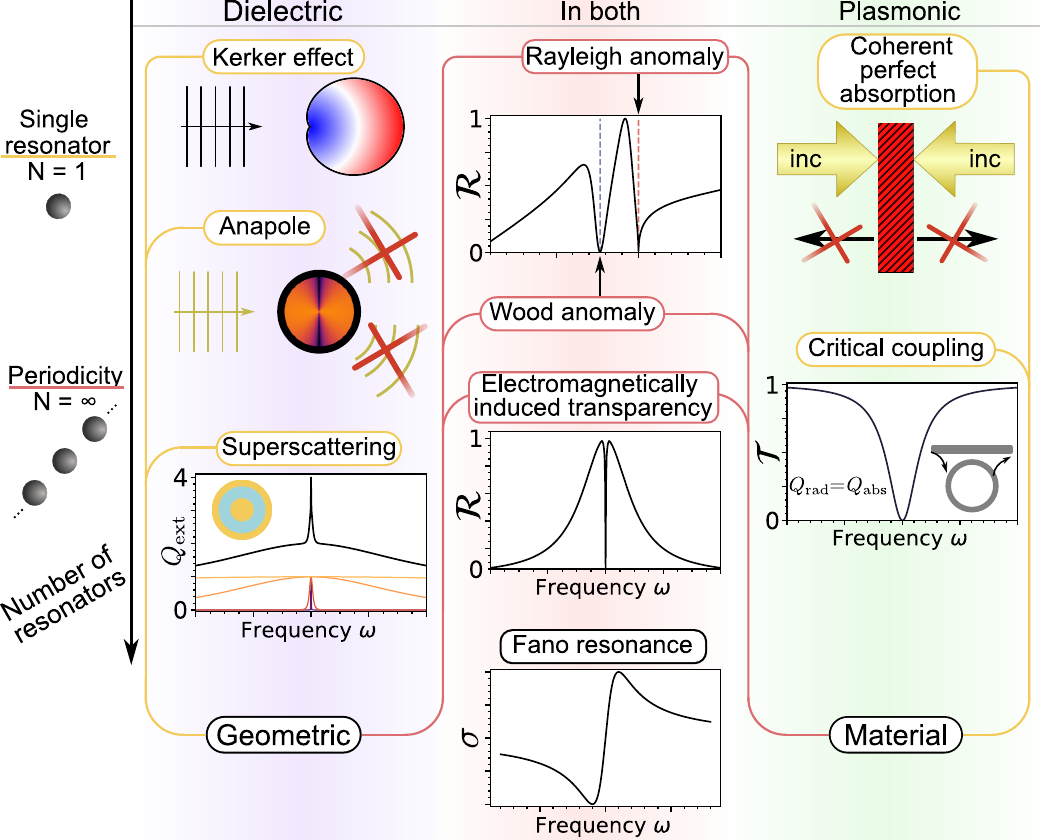}

\subsection*{Figure 2. Classification scattering effects in dielectric and plasmonic systems, organized by dimensionality of the structure.}\label{fig1_5}

Classification of scattering effects in dielectric and plasmonic systems with the notations as in Fig.~1. The scattering features are associated interference between different scattering channels or multiple resonances. The Fano resonance sketched in the bottom part does not correspond to specific system and reflects a universal wave phenomenon of resonances manifestation in scattering spectra.

\noindent\includegraphics[width=0.85\linewidth]{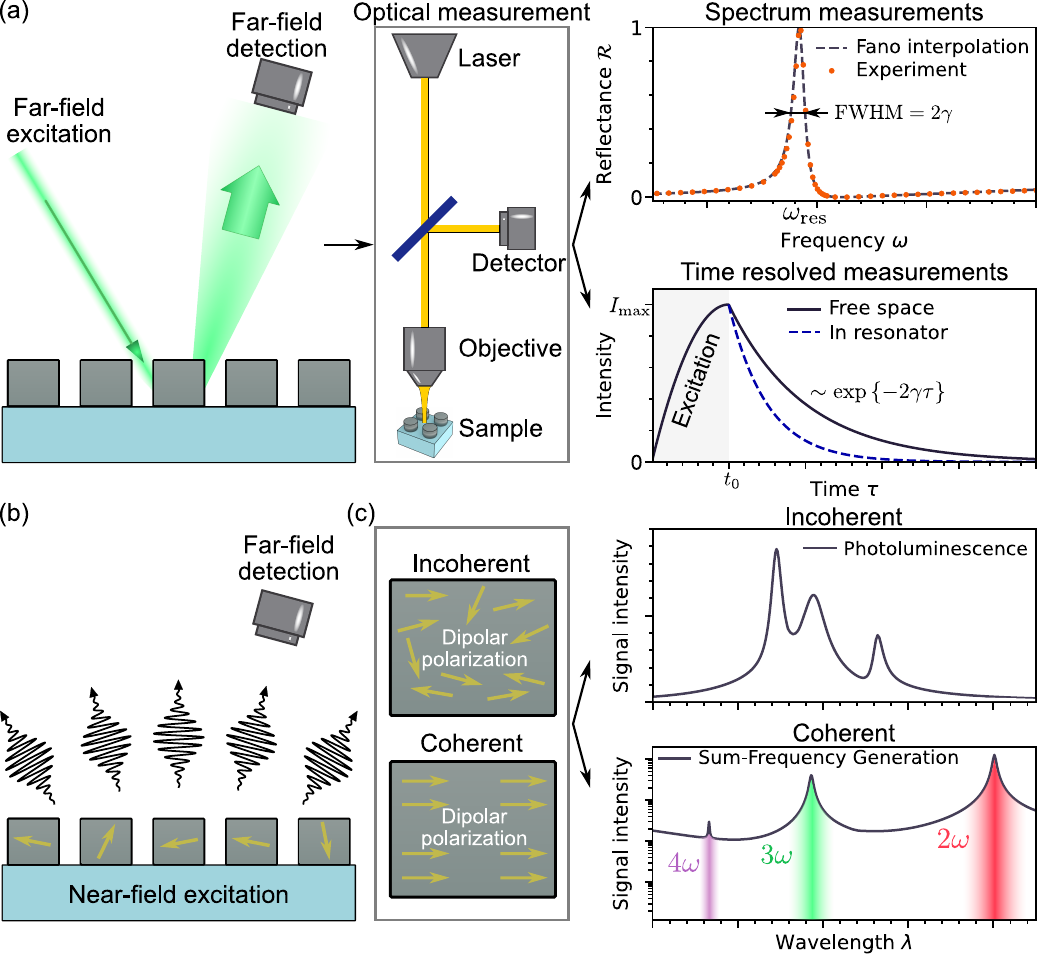}

\subsection*{Figure 3. Approaches to resonance excitation and characterization.}\label{fig:2}

(a) (Left) Far-field excitation and far-field detection setup schematic. (Right) Spectral vs. time-resolved measurement approaches.  (b) Near-field excitation and far-field detection setup schematic. (c) Incoherent (photoluminescence) and coherent (harmonic generation) near-field excitation mechanisms. Images are adapted from Refs.~\cite{fanAnalysisGuidedResonances2002,tanaka2007dynamic} (a, right), Refs.~\cite{lichtman2005fluorescence,zograf2022high} (c).

\noindent\includegraphics[width=0.85\linewidth]{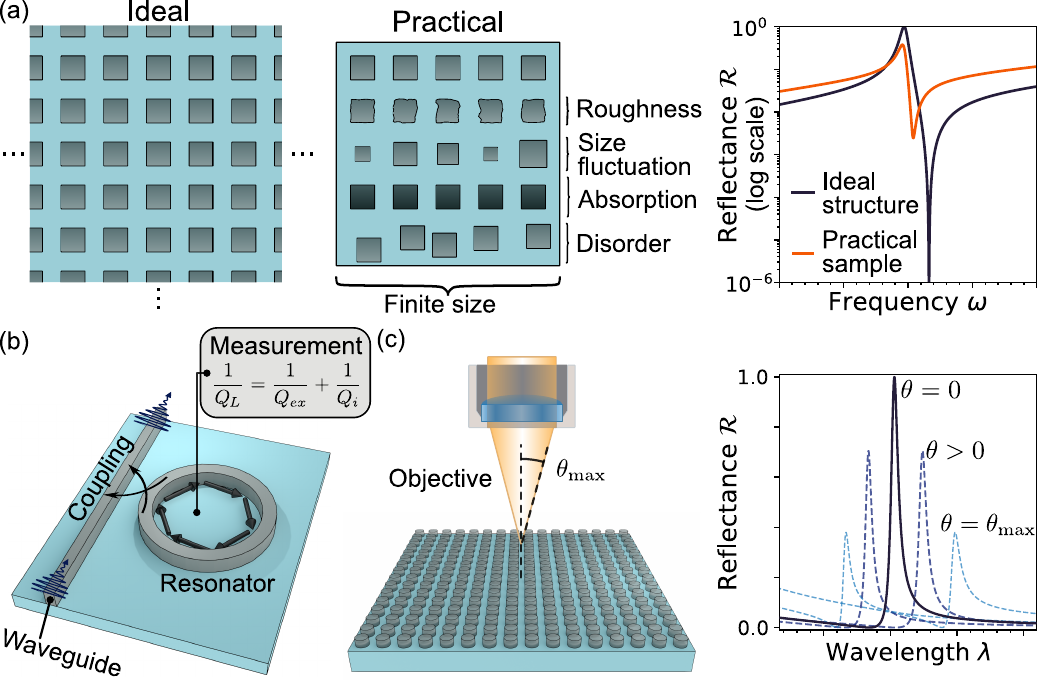}

\subsection*{Figure 4. Fabrication and measurement back-action on observable resonance parameters.}\label{fig:3}

(a) Fabrication effects. (Left) Schematic of practical practical imperfections, such as size finiteness, material roughness and absorption, fluctuation and disorder of structural elements, modifying the morphology of an ideal photonic structure. (Right) Corresponding broadening and resonance shift due to fabrication imperfections. (b, c) Measurement effects. (b) Modification of observed loaded quality factor for WGM resonator due to waveguide coupling back-action. (c) Effect of finite excitation beam divergence (numerical aperture) on the resonant linewidth and position. Images are adapted from Refs. ~\cite{bulgakov2019high,sun2021tunable,kuhne2021fabrication} (a), Ref. ~\cite{srinivasan2007mode} (b), Ref.~\cite{sun2025flatband} (c).

\newpage

\bibliography{references-1}

\bigskip

\end{document}